# Analysis of time dynamics in wind records by means of multifractal detrended fluctuation analysis and Fisher-Shannon information plane

Luciano Telesca[1]* and Michele Lovallo[2]

[1]Consiglio Nazionale delle Ricerche, Istituto di Metodologie per l'Analisi Ambientale, C.da S.Loja, 85050 Tito (PZ), Italy

[2]ARPAB, 85100 Potenza, Italy

**Abstract**

The time structure of more than 10 years of hourly wind data measured in one site in northern Italy from April 1996 to December 2007 is analysed. The data are recorded by the Sodar Rass system, which measures the speed and the direction of the wind at several heights above the ground level. To investigate the wind speed time series at seven heights above the ground level we used two different approaches: i) the Multifractal Detrended Fluctuation Analysis (MF-DFA), which permits the detection of multifractality in nonstationary series, and ii) the Fisher-Shannon (FS) information plane, which allows to discriminate dynamical features in complex time series. Our results point out to the existence of multifractal time fluctuations in wind speed and to a dependence of the results on the height of the wind sensor. Even in the FS information plane a height-dependent pattern is revealed, indicating a good agreement with the multifractality. The obtained results could contribute to a better understanding of the complex dynamics of wind phenomenon.

* Corresponding author: luciano.telesca@imaa.cnr.it

# 1. INTRODUCTION

Wind energy is the fastest growing source of energy and is going to be used worldwide for its competitive cost of production compared with other traditional means; furthermore wind energy allows to address the well-known energy resource issues and environmental problems [1]. Wind energy highly depends on wind speed; in fact, wind power is proportional to the cube of wind speed. Therefore, the analysis of wind speed records is very challenging not only for better designing more appropriately and more efficiently wind power plants (the irregular waxing and waning of wind can lead to significant mechanical stress on the gear boxes and results in substantial voltage swings at the terminals [2]), but also for better understanding the underlying dynamical mechanisms. To this aim, it is crucial to investigate the inner dynamical structure of wind speed time series. A spatial and temporal analysis of long-range dependencies in wind speed was performed in Haslett and Raftery [3]. Rehman et al. [4] analysed 10 wind speed records in Saudi Arabia and found that the Weibull distribution represents a close fit. Statistical characteristics of wind speed and diurnal variation were presented by Rehman and Halawani [5].

The characterization of the temporal fluctuations of geophysical and environmental processes has always been raising great attention for the understanding of the underlying dynamical mechanisms [6]. The standard method aiming at investigating the temporal fluctuations of a process is the power spectral density $S(f)$, which is defined in terms of the Fourier functions and describes the frequency distribution of the power. Thus, for purely random processes, which are realizations of white noise, the power spectrum is approximately flat for any frequency bands, the temporal fluctuations of the process are completely uncorrelated, any sample is completely independent of the others, and no memory phenomena exist at all. On the contrary, a power-law shape of the power spectrum, which is linear if plotted on log-log scales, indicates the presence of long-range correlated structures in the process. Such behaviour, called scaling, is typical of many geophysical and environmental processes and allows to quantify the strength of the temporal fluctuations by estimating the value of the spectral exponent, also called scaling exponent [7].

Several statistical measures could be used to gain into insight the scaling dynamics of a process by means of the estimation of the scaling exponent. But in the last decade a very effective tool, called detrended fluctuation analysis (DFA), invented by Peng et al. [8], has been extensively used for determining the scaling behaviour of signals, even if these were affected by nonstationarities with unknown origin and cause [9]. Many applications in geophysical sciences [10-15], environmental [16, 17] as well as in economics [18, 19], biology and medicine [20, 21, 22] were performed using the DFA, thus revealing its universality in being used as an effective and efficient tool for time series analysis. The DFA permits to identify scaling behaviour in monofractal series or to investigate the monofractality of a series, because it leads to the estimation of a single scaling exponent. But one scaling exponent is sufficient to completely describe a process under the hypothesis that this is monofractal. Monofractals are homogeneous objects, in the sense that they have the same scaling properties, characterized by a single singularity exponent [23]. The need for more than one scaling exponent can derive from the existence of a crossover timescale, which separates regimes with different scaling behaviors, suggesting e.g. different types of correlations at small and large timescales, thus leading to different types of time dynamics intrinsic in the process [24-27]. Different values of the same scaling exponent could be required for different segments of the same series, indicating a time variation of the scaling behaviour, relying to a time variation of the underlying dynamics [28]. Furthermore, different scaling exponents can be revealed for many interwoven fractal subsets of the signal; in this case the process is not a monofractal but multifractal. A multifractal object requires many indices to characterize its scaling properties. Multifractals can be decomposed into many-possibly infinitely many-sub-sets characterized by different scaling exponents. Thus multifractals are intrinsically more complex and inhomogeneous than monofractals and characterize systems featured by a spiky dynamics, with sudden and intense bursts of high frequency fluctuations [29]. Taking into account of the independence of nonstationarity revealed by the DFA, its generalization into the multifractal detrended fluctuation analysis (MF-DFA) was developed by Kantelhardt et al. [9]. This method is, thus, able to reliably determine the multifractal scaling behavior of nonstationary series.

Up to our knowledge the first paper that investigated the multifractality in wind speed was by Kavasseri and Nagarajan [30]. They analysed four hourly averaged wind speed records in North Dakota at a height of 20m above the ground level. They found that the binomial cascade multiplicative model could represent a close fit to the data, although spatial and temporal variations in wind speed are influenced by pressure gradients, turbulence, temperature and topography.

## 2. WIND DATA

We analysed the time series of the hourly speed of wind measured by a Sodar Rass system in northern Italy from April 1996 to December 2007. The data are available free of charge on the following internet website http://www.istitutoveneto.it/venezia/dati/atmosfera/dati_enel/234sidfg45.htm. Such system allows to measure the speed of wind at different heights from the ground. We analysed the time series of wind speed at height H=50m, 77m, 104m, 131m, 158m, 186m and 213m above the ground level. Fig.1 shows the seven time series of wind speed.

## 3. METHODS AND DATA ANALYSIS

### 3.1. Multifractal detrended fluctuation analysis

The main features of multifractals is to be characterized by high variability on a wide range of temporal scales, associated to intermittent fluctuations and long-range power-law correlations.

The data examined in this paper present clear irregular dynamics (Fig. 1), characterized by sudden bursts of high frequency fluctuations, which suggest to perform a multifractal analysis, thus evidencing the presence of different scaling behaviours for different intensities of fluctuations. Furthermore, the signal appear nonstationary, and, for this reason, we applied the Multifractal Detrended Fluctuation Analysis (MF-DFA), which operates on the time series $x(i)$, where $i=1,2,...,N$ and $N$ is the length of the series. With $x_{ave}$ we indicate the mean value

$$x_{\text{ave}} = \frac{1}{N}\sum_{k=1}^{N} x(k) \ . \quad (1)$$

We assume that *x(i)* are increments of a random walk process around the average $x_{ave}$, thus the "trajectory" or "profile" is given by the integration of the signal

$$y(i) = \sum_{k=1}^{i} [x(k) - x_{ave}] \quad . \quad (2)$$

Furthermore, the integration will reduce the level of measurement noise present in observational and finite records. Next, the integrated time series is divided into nonoverlapping $N_S = int(N/s)$ segments of equal length *s*. Since the length *N* of the series is often not a multiple of the considered time scale *s*, a short part at the end of the profile *y(i)* may remain. In order not to disregard this part of the series, the same procedure is repeated starting from the opposite end. Thereby, $2N_S$ segments are obtained altogether. Then we calculate the local trend for each of the $2N_S$ segments by a least square fit of the series. Then we determine the variance

$$F^2(s,\nu) = \frac{1}{s} \sum_{i=1}^{s} \{y[(\nu-1)s+i] - y_\nu(i)\}^2 \quad (3)$$

for each segment $\nu$, $\nu = 1,..,N_S$ and

$$F^2(s,\nu) = \frac{1}{s} \sum_{i=1}^{s} \{y[N-(\nu-N_S)s+i] - y_\nu(i)\}^2 \quad (4)$$

for $\nu = N_S+1,..,2N_S$. Here, $y_\nu(i)$ is the fitting line in segment $\nu$. Then, after detrending the series, we average over all segments to obtain the *q*-th order fluctuation function

$$F_q(s) = \left\{ \frac{1}{2N_S} \sum_{\nu=1}^{2N_S} [F^2(s,\nu)]^{\frac{q}{2}} \right\}^{\frac{1}{q}} \quad (5)$$

where, in general, the index variable *q* can take any real value except zero. Repeating the procedure described above, for several time scales *s*, $F_q(s)$ will increase with increasing *s*. Then analyzing log-log plots $F_q(s)$ versus *s* for each value of $q$, we determine the scaling behaviour of the fluctuation functions. If the series $x_i$ is long-range power-law correlated, $F_q(s)$ increases for large values of $s$ as a power-law

$$F_q(s) \propto s^{h(q)}. \quad (6)$$

Monofractal time series are characterized by $h(q)$ independent of $q$. The different scaling of small and large fluctuations will yield a significant dependence of $h(q)$ on $q$. For positive $q$, the segments $v$ with large variance (i.e. large deviation from the corresponding fit) will dominate the average $F_q(s)$. Therefore, if $q$ is positive, $h(q)$ describes the scaling behaviour of the segments with large fluctuations; and generally, large fluctuations are characterized by a smaller scaling exponent $h(q)$ for multifractal time series. For negative $q$, the segments $v$ with small variance will dominate the average $F_q(s)$. Thus, for negative $q$ values, the scaling exponent $h(q)$ describes the scaling behaviour of segments with small fluctuations, usually characterized by a larger scaling exponents.

The value $h(0)$ corresponds to the limit $h(q)$ for $q \to 0$, and cannot be determined directly using the averaging procedure of Eq. 5 because of the diverging exponent. Instead, a logarithmic averaging procedure has to be employed,

$$F_0(s) \equiv \exp\left\{\frac{1}{4N_S}\sum_{v=1}^{2N_S}\ln[F^2(s,v)]\right\} \approx s^{h(0)} \quad . \quad (7)$$

In general the exponent $h(q)$ will depend on $q$. For stationary time series, $h(2)$ is the well defined Hurst exponent $Hu$ [31]. Thus, we call $h(q)$ the generalized Hurst exponent. Fig. 2 shows the fluctuation functions $F_2(s)$ for the seven wind speed time series. It is striking that whatever the height of measurement, the fluctuation functions increase as a power law up to the crossover $s_c$~4-6 months, after which the fluctuation functions tend to flatten, thus indicating the presence of two different dynamics. It may be very likely that the origin of such crossover is seasonal and linked with meteo-climatic phenomena. Even changing the order of the moment q, the crossover still remains, as shown in Fig. 3, which shows, as an example, the fluctuation functions for $q$=-10, 0, +10 for the height $H$=104m. This indicates that the seasonal crossover, which approximately separates two scaling regions, and, thus, discriminates two different dynamics for timescales respectively lower and higher than $s_c$, is a characteristic parameter of the wind data, and does not depend on the range of variability of the data, because it is present for large ($q$>0) as well as small ($q$<0) fluctuations. Furthermore the different slopes of the fluctuation curves (~0.69 for $q$=+10, ~0.83 for $q$=0 and ~0.95 for $q$=-10) indicate that small and large wind fluctuations scale differently, indicating the presence of multifractal dynamics.

Fig. 4 shows the $q$-dependence of the generalized Hurst exponent $h(q)$ determined by fits in the regime 30 hours<$s$<$10^{3.5}$ hours for the seven wind time series (the superior limit was chosen to be less than the crossover $s_c$). Although the $h(q)$~$q$ relationships are quite similar, slight difference in the multifractal degree can be evaluated. To this aim, the computation of the multifractal spectrum by means of the Legendre transform was performed. The multifractal scaling exponents $h(q)$ defined in Eq. 6 are directly related to the scaling exponents $\tau(q)$ defined by the standard partition function multifractal formalism [9]. Suppose that the series $x_k$ is a stationary and normalized sequence. Then the detrending procedure of the MF-DFA is not required.

Thus the DFA can be replaced by the Fluctuation Analysis (FA), for which the variance is defined as

$$F_{FA}^2(s,\nu) = [y(\nu s) - y((\nu-1)s)]^2. \quad (8)$$

Inserting this definition in Eq. 5 and using Eq. 7, we obtain

$$\left\{\frac{1}{2N_S}\sum_{\nu=1}^{2N_S}[F_{FA}^2(s,\nu)]^q\right\}^{\frac{1}{q}} \approx s^{h(q)} \quad (9)$$

For sake of simplicity, we assume that the length $N$ of the sequence is a integer multiple of the scale $s$, obtaining $N_S=N/s$ and therefore

$$\sum_{\nu=1}^{N_S}[F_{FA}^2(s,\nu)]^q \approx s^{qh(q)-1}. \quad (10)$$

The term $[y(\nu s)-y((\nu-1)s)]$ is the sum of the numbers $x_k$ within each segment $\nu$ of size $s$. This sum is known as the box probability $p_s(\nu)$ in the standard multifractal formalism for normalized series $x_k$. The scaling exponent $\tau(q)$ is usually defined via the partition function $Z_q(s)$,

$$Z_q(s) = \sum_{\nu=1}^{N_S}|p_s(\nu)|^q \approx s^{\tau(q)}, \quad (11)$$

where $q$ is a real parameter as in the MF-DFA. Eq. 11 is identical to Eq. 10, therefore

$$\tau(q)=qh(q)-1. \quad (12)$$

In this equation $h(q)$ is different from the generalized multifractal dimensions $D(q) = \frac{\tau(q)}{q-1}$; in fact, while $h(q)$ is independent of $q$ for a monofractal series, $D(q)$ depends on $q$ in that case [19]. Therefore, monofractal series with long-range correlations are characterized by linearly dependent $q$-order exponent $\tau(q)$, i.e. the exponents $\tau(q)$ of different moments $q$ are linearly dependent on $q$

$$\tau(q)=Hq-1, \quad (13)$$

with a single Hurst exponent,

$$H=d\tau/dq=\text{const}. \quad (14)$$

Long-range correlated multifractal signals have a multiple Hurst exponent, i.e. the generalized Hurst exponent $h(q)$,

$$h(q)=d\tau/dq \neq \text{const}, \quad (15)$$

where $\tau(q)$ depends nonlinearly on $q$ [32].

The singularity spectrum $f(\alpha)$ is related to $\tau(q)$ by means of the Legendre transform,

$$\alpha = \frac{d\tau}{dq} \quad (16)$$

$$f(\alpha)=q\alpha-\tau(q), \quad (17)$$

where $\alpha$ is the Hölder exponent and $f(\alpha)$ indicates the dimension of the subset of the series that is characterized by $\alpha$. The singularity spectrum quantifies in details the long-range correlation properties of a time series. Fig. 5 shows the multifractal spectrum $f(\alpha)$ for the seven wind time series. The multifractal spectrum gives information about the relative importance of various fractal exponents present in the series. In particular the width of the spectrum indicates the range of present exponents. To be able to make quantitative characterization of multifractal spectra, the spectrum was fitted to a quadratic function [33] around the position of its maximum at $\alpha_0$, i.e. $f(\alpha) = A(\alpha-\alpha_0)^2 + B(\alpha-\alpha_0) + C$ : the coefficients are obtained by an ordinary least-square procedure. In this fitting the additive constant $C=f(\alpha_0)=1$.

Parameter *B* serves as an asymmetry parameter, which is zero for symmetric shapes, positive or negative for a left- or right-skewed (centered) shape, respectively. To obtain an estimate of the range of possible fractal exponents, we measured the width of the spectrum, extrapolating the fitted curve to zero. The width of the spectrum was then defined as $W = \alpha_1 - \alpha_2$, with $f(\alpha_1) = f(\alpha_2) = 0$. Obviously, negative values for $f(\alpha)$ are permitted, and the proposed measure of the width of the multifractal spectrum is equivalent to the calculation of the range (maximum-minimum) of the $h(q)$ scaling exponents [32]. The measure of the width of multifractal spectra as a measure of degree of multifractality has also been suggested in relation to the investigation of past climate variations [34]. With low $\alpha_0$, the process becomes correlated, for example if the process has had the tendency to move upward in the past, it will move upward with a probability larger than 1/2 in the next time step. Roughly speaking, a small value of $\alpha_0$ means that the underlying process is more regular in appearance. The width of the spectrum W is a measure of how wide the range of fractal exponents found in the signal is; and, thus, it measures the degree of multifractality of the series. The wider the range of possible fractal exponents, the "richer" is the process in structure. Finally, the asymmetry parameter *B* captures the dominance of low or high fractal exponents with respect to the other. A right-skewed spectrum indicates relatively strongly weighted low fractal exponents, and high ones for left-skewed shapes. Fig. 6 shows the multifractal parameters (maximum, asymmetry and width) inferred from the multifractal spetrum of each wind speed time series. It is clear a pattern of the multifractal parameters with the height: at large heights wind speed appears relatively more correlated and regular (relatively low values for the maximum $\alpha_0$) with a relative dominance of low fractal exponents (negative asymmetry *B*) and a lower degree of multifractality (lower value of width), thus suggesting more homogenous underlying dynamics. In particular the values of the multifractal width are in very good agreement with those obtained by Kavasseri and Nagarajan applying the binomial multiplicative cascade model [30].

In order to understand what is the type of multifractality underlying the q-dependence of the generalized Hurst exponent, we applied the random shuffle method to generate 100 surrogate series for each wind speed time series. Generally, two different types of multifractality in time series can be discriminated: (i) due to a broad probability density function, and (ii) due to different long-range correlations for small and large fluctuations. In the shuffling procedure the values are put into random order, and although all correlations are destroyed, the probability density function remains unchanged. Hence the shuffled series coming from multifractals of type (ii) will exhibit simple random behaviour with $h_{shuf}(q)=0.5$. While those coming from multifractals of type (i) will show $h(q)=h_{shuf}(q)$, since the multifractality depends on the probability density [9]. If both types of multifractality characterize the time series, thus the shuffled series will show weaker multifractality than the original one. Fig. 7 shows the results of the generalized Hurst exponents versus q, averaged over 100 randomly shuffled versions of the original time series. The error bars delimit the 1-$\sigma$ range around the mean values. The mean $h_{shuf}(q)$-values range around 0.5 for any height, but with a slight *q*-dependence; this indicates that the most multifractality of the wind speed data is due to different long-range correlations for small and large fluctuations. Fig. 8 shows the multifracatal spectra, through the Legendre transform, of the shuffles. For comparison the multifractal spectrum of the original wind speed series at $H$=50m is also shown. It is clearly evident that the shuffled series are characterized by lower multifractality degree. And this confirms that the most mutlifractality of the wind series depends on the different long-range correlation properties for small and large fluctuations.

### 3.2 Fisher-Shannon method

A further approach used to investigate the dynamics of the wind speed records is the Fisher-Shannon (FS) information plane. The Fisher Information Measure (FIM) is a powerful tool to investigate complex and nonstationary signals; the Shannon entropy is the well-known magnitude to quantify the degree of disorder in dynamical systems. The FIM was introduced by Fisher in 1925 in the context of statistical estimation [35]. In a seminal paper Frieden [36] has shown that FIM is a versatile tool to describe the evolution laws of physical systems. FIM allows to accurately describe the behavior of dynamic systems, and to characterize the complex signals generated by these systems [37]. This approach has been used by Martin et al. to characterize the dynamics of EEG signals [38]. Martin et al. [39] have shown the informative content of FIM in detecting significant changes in the behavior of nonlinear dynamical systems disclosing, thus, FIM as an important quantity involved in many aspects of the theoretical and observational description of natural phenomena. The FIM was used in studying several geophysical and environmental phenomena, revealing its ability in describing the complexity of a system [40, 41] and suggesting its use as to reveal reliable precursors of critical events [42, 43].

The Shannon entropy is a measure of the amount of information in a certain information source and represents the degree of indeterminacy in a certain system [44]. The Shannon entropy can be used to define the degree of uncertainty involved in predicting the output of a probabilistic event [45]. For discrete distributions, this means that if one predicts the outcome exactly before it happens, the probability will be a maximum value and, as a result, the Shannon entropy will be a minimum. If one is absolutely able to predict the outcome of an event, the Shannon entropy will be zero. Such is not the case for distributions (probability densities) on a continuous variable, ranging e.g. over the real line. In this case, the Shannon entropy can reach any arbitrary value, positive or negative. Therefore, the use of the power entropy (that is defined below) avoids the difficulty of dealing with negative information measures. Shannon entropy provides a scientific method to understand the essential state of things [46, 47].

Let us introduce the relevant Fisher- and Shannon-associated quantities [39]. Let $f \equiv q^2$ be a probability density in $\Re^d$ ($d \geq 1$). Fisher's quantity of information associated to $f$ (or to the probability amplitude q) is defined as the (possibly infinite) non-negative number $I$

$$I(f) = \int_{\Re^d} d\mathbf{x} \frac{|\nabla f|^2}{f} \quad (18)$$

or in terms of the amplitudes

$$I(q) = \int_{\Re^d} d\mathbf{x}(\nabla q \cdot \nabla q). \quad (19)$$

where $\nabla$ is the differential operator. This formula defines a convex, isotropic functional $I$, which was first used by Fisher [35] for statistical purposes. It is clear from Eq. 19 that the integrand, being the scalar product of two vectors, is independent of the reference frame [39].

Let us focus the attention on the one-dimensional case. Let us consider a measurement $x$ whose probability density function is denoted as $f(x)$. Its FIM is defined as

$$I = \int_{-\infty}^{+\infty} \left(\frac{\partial}{\partial x} f(x)\right)^2 \frac{dx}{f(x)}. \quad (20)$$

The Shannon entropy is given by the following formula [37]:

$$H_X = -\int_{-\infty}^{+\infty} f_X(x) \log f_X(x) dx \quad (21).$$

For convenience the alternative notion of entropy power [48]

$$N_X = \frac{1}{2\pi e} e^{2H_X} \quad (22)$$

will be used rather than the entropy $H_X$. The use of the power entropy $N_X$ instead of the Shannon one $H_X$ arises from the so-called 'isoperimetric inequality' [48-51], a lower bound to the Fisher-Shannon product which reads as $IN_X \geq d$, where $d$ is the dimension of the space. The 'isoperimetric inequality' suggests that the FIM and the Shannon entropy are intrinsically linked, so that the dynamical characterization of signals should be improved when analyzing them in the so called Fisher-Shannon (FS) information plane [37], in which the $y$- and $x$-axis are the FIM and the Shannon entropy (as outlined above, instead of the Shannon entropy we will use the entropy power $N_X$). Vignat and Bercher [37] showed that the simultaneous examination of both Shannon entropy and FIM through the FS plane could improve the characterization of the non-stationary behavior of complex signals. The product $IN_X$ can be considered as a statistical measure of complexity [49]. The line $IN_X=1$ separates the FS plane in two parts: one allowed ($IN_X>1$) and one not allowed ($IN_X<1$), and the distance of a signal point from the 'isocomplexity line' $IN_X=1$ can measure the degree of complexity of the signal.

Eqs. 20 and 21 involve the calculation of the probability density function (pdf) $f(x)$.

An estimation of the pdf $f(x)$ may be obtained by means of the kernel density estimator technique [52, 53]. The kernel density estimator provides an approximate value of the density in the form

$$\hat{f}_M(x) = \frac{1}{Mb} \sum_{i=1}^{M} K\left(\frac{x - x_i}{b}\right). \quad (23)$$

where $M$ is the number of data and $K(u)$ is the kernel function, which is a continuous non-negative and symmetric function satisfying

$$K(u) \geq 0 \text{ and } \int_{-\infty}^{+\infty} K(u)du = 1, \quad (24)$$

whereas *b* is the bandwidth. In our estimation procedure the kernel used is the Gaussian of zero mean and unit variance. In this case

$$\hat{f}_M(x) = \frac{1}{M\sqrt{2\pi b^2}} \sum_{i=1}^{M} e^{-\frac{(x-x_i)^2}{2b^2}}. \quad (25)$$

The Gaussian kernel allows to evaluate the kernel density estimator and the bandwidth with a low computational complexity [54].

Fig. 9 shows the Fisher-Shannon information plane for wind speed data measured in northern Italy. Each symbol represents a record of the wind speed at a certain height over the ground level. It is clearly visible a pattern with the height, revealing that the FIM of wind speed (and correspondingly the Shannon entropy) decreases (increases) with the height. From these results it can be deduced that the content of information or the degree of order is higher at lower heights above the ground level.

## 4. CONCLUSIONS

The mechanisms underlying wind dynamics are complex. The multifractal analysis and the Fisher-Shannon approach, performed in the present study, have led to a better description of such complexity. Although the physical interpretation of such results is not a simple task, it is noteworthy that the both the analysis (MF-DFA and FS) have furnished very consistent results, indicating a height-dependent behavioural trend in wind speed. Further similar analysis performed over different wind speed time series, measured in different sites and different periods could gain into insight a better understanding of the complexity of wind phenomena.

**Figure captions**

Fig. 1. Hourly wind speed time series at 50m (a), 77m (b), 104m (c), 131m (d), 158m (e), 186m (f), 213m (g) above the ground level.

Fig. 2. $F_2(s)$~s curves for the wind speed time series plotted in Fig. 1.

Fig. 3. Fluctuation functions for q=-10, 0, +10 for the height H=104m.

Fig. 4. h(q)-q relationships for the wind speed time series, calculated on the range scales 30 hours<s<$10^{3.5}$ hours.

Fig. 5. Legendre spectra.

Fig. 6. Multifractal parameters: a) maximum, b) asymmetry, c) width.

Fig. 7. h(q)-q relationships averaged over 100 shuffled wind speed time series.

Fig. 8. Comparison between the Legendre spectra of the shuffled wind speed time series and the that of the original wind speed at 50 m above the ground level.

Fig. Fisher-Shannon information plane for the wind speed time series plotted in Fig. 1.

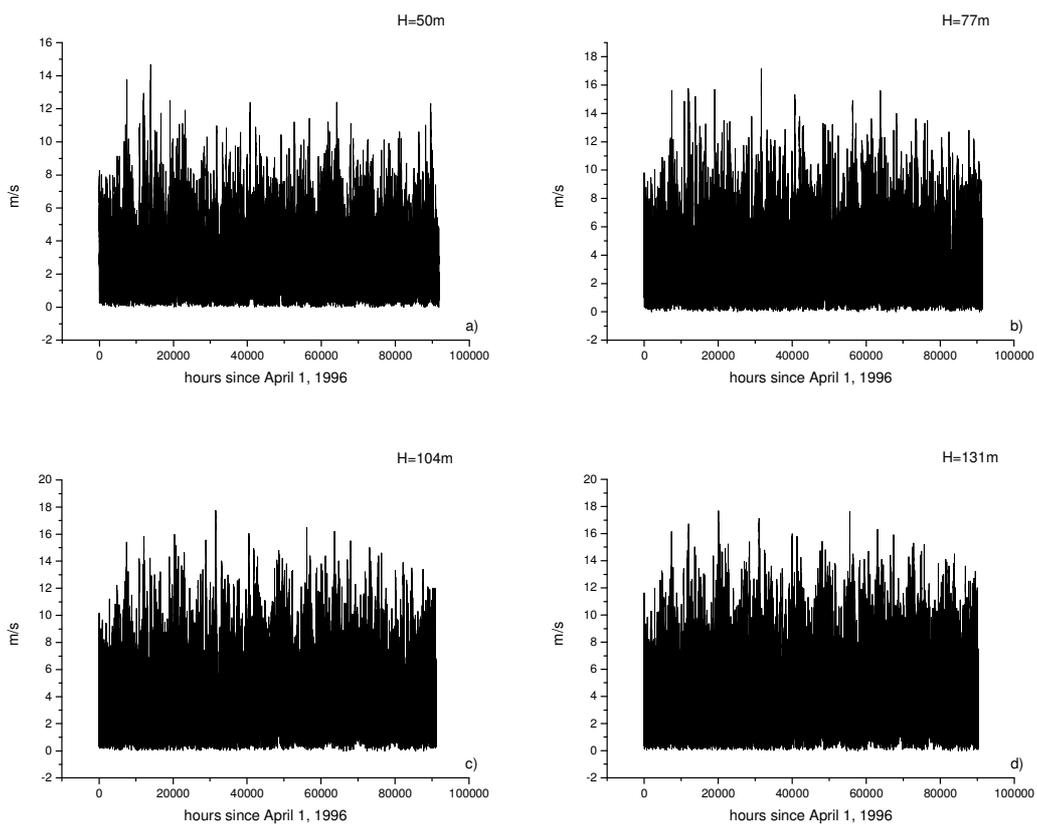

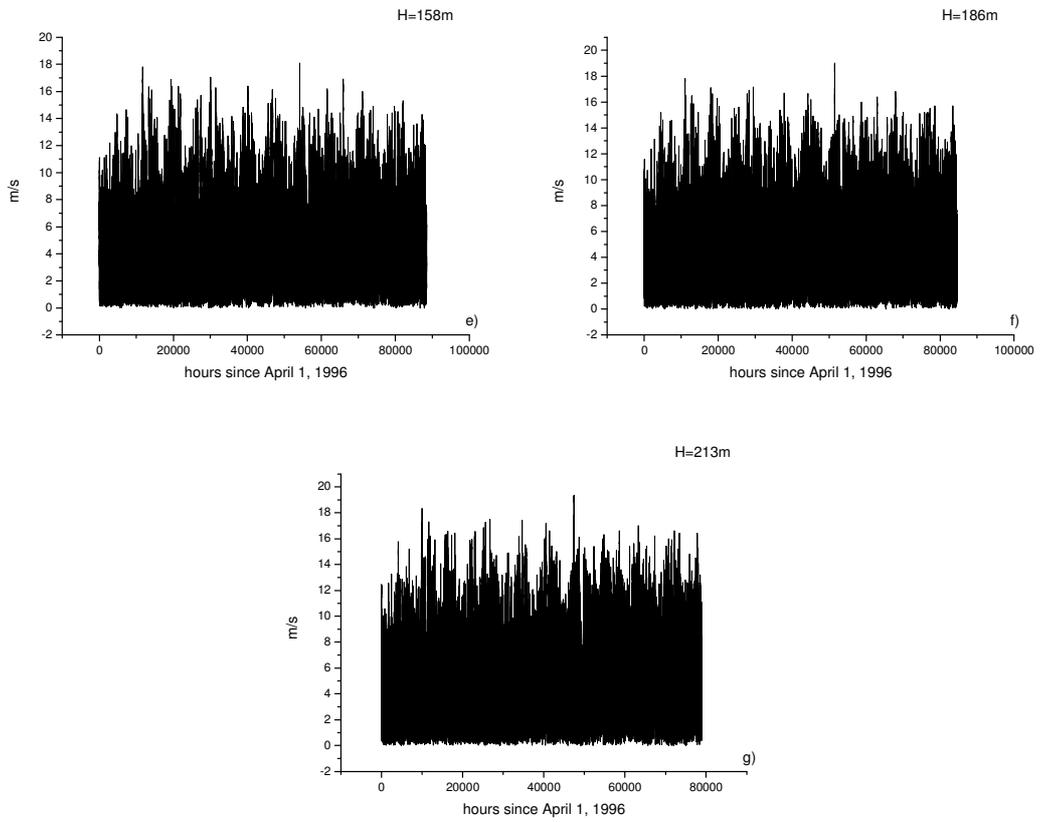

Fig. 1

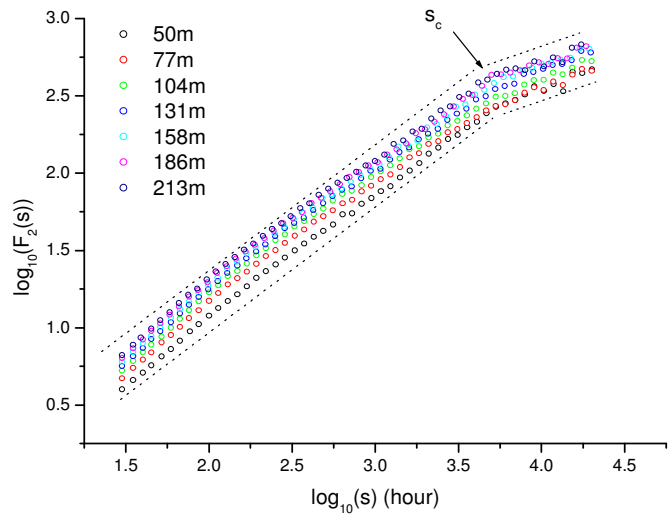

Fig. 2

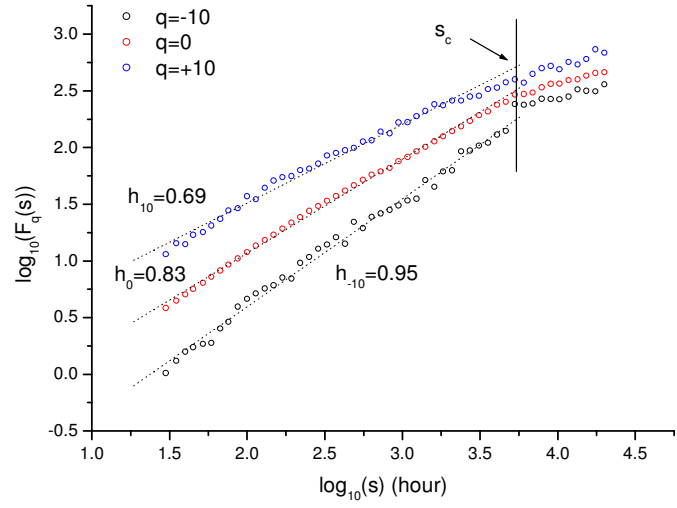

Fig. 3

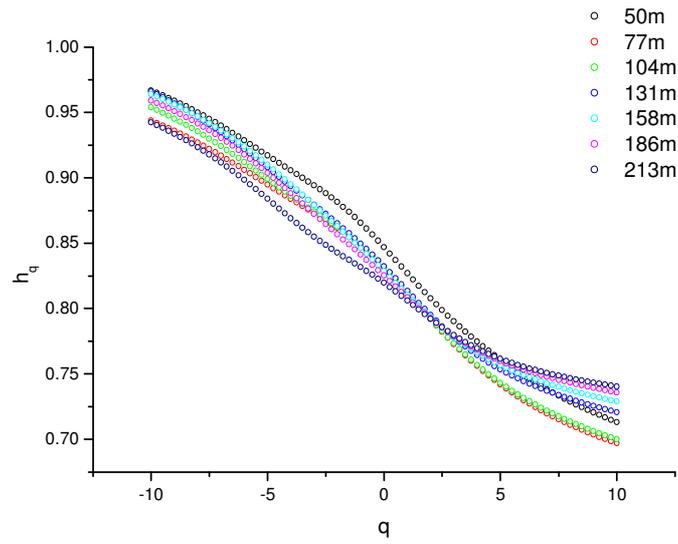

Fig. 4

28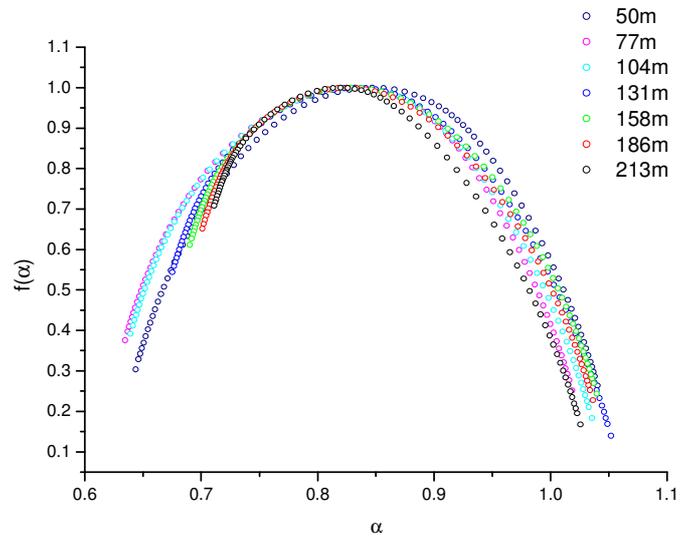

Fig. 5

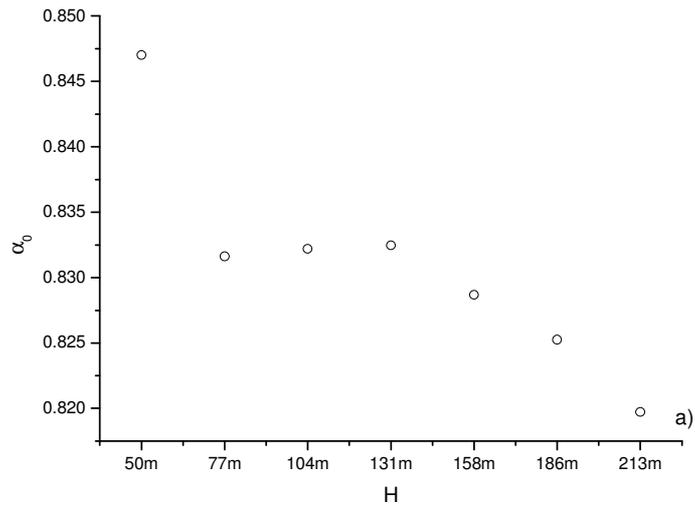

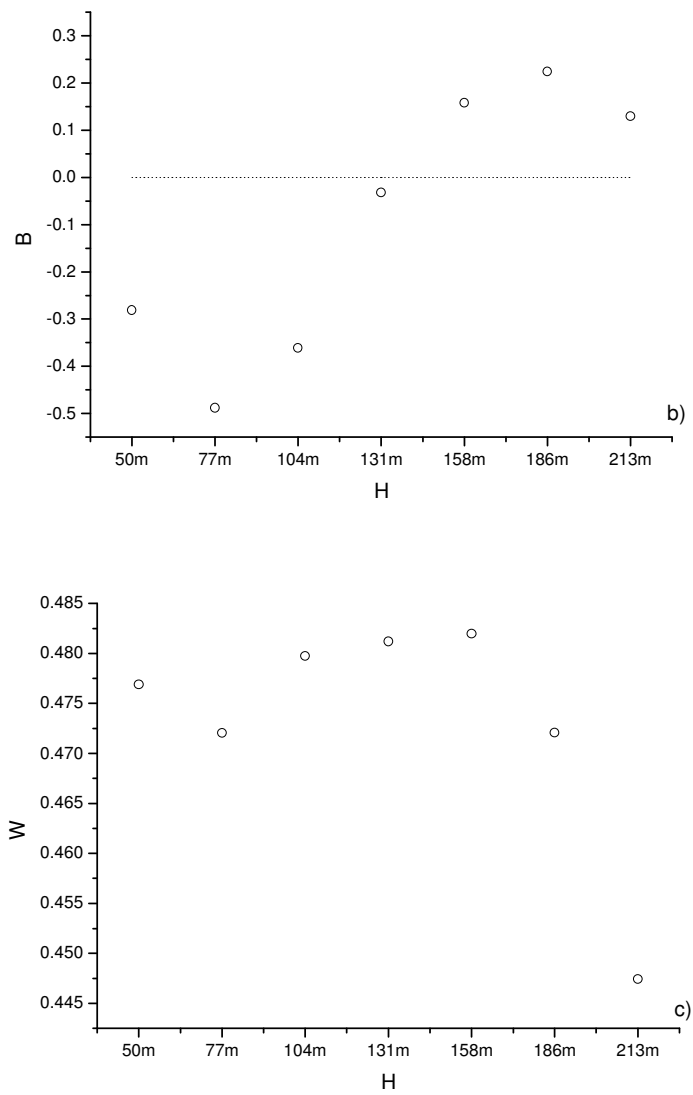

Fig. 6

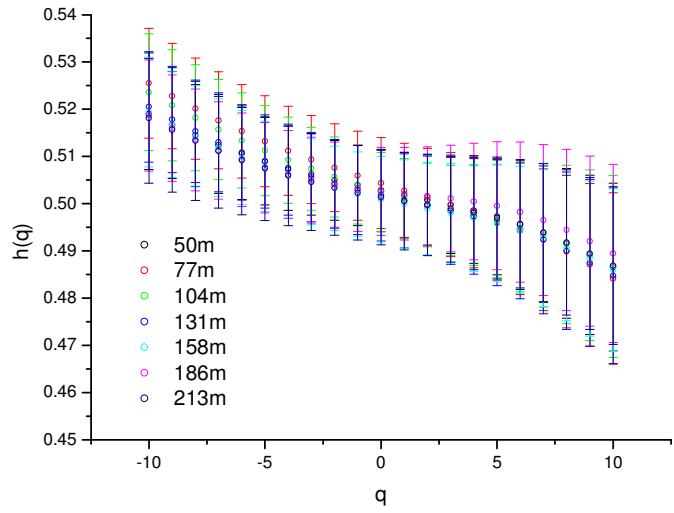

Fig. 7

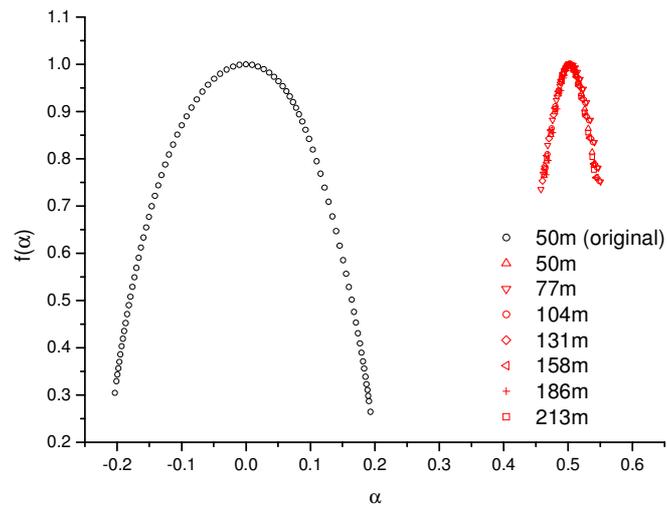

Fig. 8

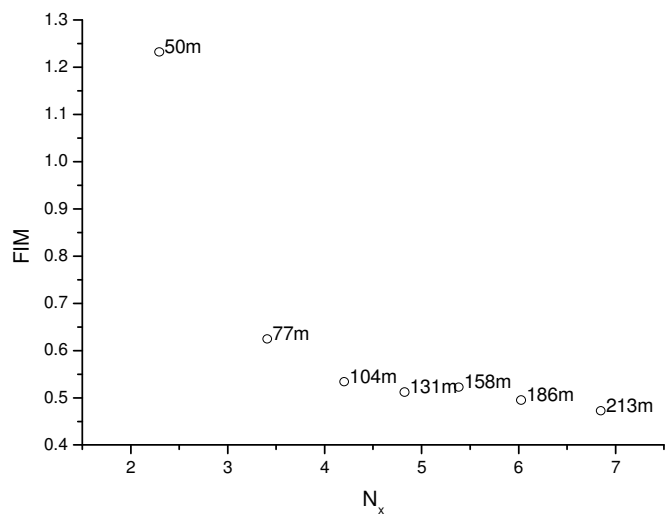

Fig. 9